\documentclass[11pt,a4paper]{article}
\usepackage{graphicx}

\title{On the stability of steady general-relativistic accretion and analogue black holes}
\author{Patryk Mach \vspace{1ex}\\
\normalsize\textit{M. Smoluchowski Institute of Physics, Jagiellonian University},\\
\normalsize\textit{Reymonta 4, 30-059 Krak\'ow, Poland}\\
\normalsize{e-mail: }\normalsize\texttt{mach@th.if.uj.edu.pl}}
\date{}

\begin{document}

\maketitle

\begin{abstract}
Investigation of general-relativistic spherically symmetric steady accretion of self-gravitating perfect fluid onto compact objects reveals the existence of two weakly accreting regimes. In the first (corresponding to the test fluid approximation) the mass of the central object is much larger than the mass of the accreting fluid; in the second the mass of the fluid dominates. The stability of the solutions belonging to the first regime has been proved by Moncrief. In this work we report the results of a series of numerical studies demonstrating stability of massive solutions, i.e., belonging to the second of the aforementioned regimes. It is also shown that a formal analogy between ``sonic horizons'' in the accretion picture and event horizons in general relativity is rather limited. The notion of a ``sonic horizon'' is only valid in a linear regime of small hydrodynamical perturbations. Strong perturbations can still escape from beneath the ``sonic horizon.''

{\vspace{1em}\hspace{-\parindent}\textbf{Key words:} general-relativistic hydrodynamics, accretion.}

\end{abstract}

\section{Introduction}

The model of spherically symmetric self-gravitating accretion in general-relativistic hydrodynamics has been studied by Malec~\cite{malec} and Karkowski et al.~\cite{kkmms}. It describes a spherically symmetric compact cloud of gas that, due to gravity, falls onto a central body (a star or a black hole).

The underlying mathematical problem consists of solving Einstein equations for spherically symmetric space time, where the energy-momentum tensor is that of perfect fluid, and where a suitable assumption of quasi-stationarity is made --- one searches for a solution for which quantities such as the density, the pressure, the radial velocity etc., computed at a given areal radius, are constant in time. In the accretion case such conditions can be satisfied only approximately, as due to the inflow of gas the mass of the central object increases and this must be reflected in the change of other quantities (most obviously the space-time metric). Thus, the solutions of particular interest are those for which the accretion rate is sufficiently small, so that for a long time (as compared to the dynamical scale of the system) the central mass remains almost unchanged and the assumption of quasi-stationarity can be justified a posteriori.

It was shown in \cite{kkmms} that for specified boundary conditions such as the total mass of the system (the sum of the mass of the central body and the mass of the fluid), the areal radius of the accreting cloud and the local speed of sound at the outer boundary, there exist two branches of solutions that correspond to small accretion rates. Solutions belonging to the first branch can be approximated by well known test-fluid solutions (cf.\,\cite{bondi} for the Newtonian model and \cite{michel} for a general-relativistic version). In this case the mass of the central body is much larger than the mass of the accreting fluid and the motion occurs practically in the fixed Schwarzschild space-time. The other branch consists of solutions for which the mass of the fluid is larger than the mass of the central object. In order to obtain these solutions one has to solve the complete set of Einstein equations.

Stability of the solutions in the test-fluid approximation was proved by Moncrief~\cite{moncrief} for linearized perturbations. Unfortunately the technique of the proof breaks when applied to self-gravitating flows. We have, therefore, performed a series of numerical studies of the stability of solutions belonging to the massive branch in the full, nonlinear regime. In each examined case the simulated flow was stable.

\section{Stationary solutions}

We will consider a spherically symmetric cloud of gas that falls onto a compact central object. The general, spherically symmetric metric can be described by the line element of the form
\begin{equation}
\label{comoving_metric}
d s^2 = -N^2 d {\tilde t}^2 + A d r^2 + R^2 \left(d\theta^2 + \sin^2 \theta d\phi^2 \right),
\end{equation}
where $N$, $A$ and $R$ are functions of the coordinates $\tilde t$ and $r$. Let us further consider a foliation of such a space-time by slices of a constant time $\tilde t$ and a two dimensional sphere of a constant coordinate radius $r$ embedded in a given temporal slice. The trace of the extrinsic curvature of such a sphere (twice the mean curvature) reads $k = 2 \partial_r R / \left( R \sqrt{A} \right)$.

The accreting gas is described by the energy-momentum tensor of perfect fluid
\begin{equation}
\label{energy_momentum_tensor}
T^{\mu\nu} = \left(p + \rho \right) u^\mu u^\nu + p g^{\mu \nu}.
\end{equation}
Here $p$ denotes the pressure, $\rho$ is the energy density and $u^\mu$ are the components of the four-velocity of the fluid. The metric tensor is denoted with $g^{\mu \nu}$.

Let $\tilde t$, $r$, $\theta$, and $\phi$ appearing in (\ref{comoving_metric}) be the comoving coordinates. The comoving gauge can be obtained by imposing a suitable geometric condition expressed in terms of the extrinsic curvature of time slices \cite{malec}. Then, assuming (\ref{energy_momentum_tensor}), one can show that $u^r = u^\theta = u^\phi = 0$.

Let us also introduce a function $U = \partial_{\tilde t} R/N$. It plays a role of the radial velocity in the coordinate system obtained by the transformation $(\tilde t, r) \mapsto (t^\prime = \tilde t, r^\prime = R(\tilde t,r))$. In these coordinates the components of the four-velocity read
\[ u^{t^\prime} = \frac{1}{N}, \;\;\; u^R = U, \;\;\; u^\theta = u^\phi = 0. \]

Another quantity that we introduce is the quasi-local mass
\[ m(R) = m_\mathrm{tot} - 4 \pi \int_R^{R_\infty} {R^\prime}^2 \rho d R^\prime. \]
Here $m_\mathrm{tot}$ denotes the total (asymptotic) mass of the considered configuration, and $R_\infty$ is the areal radius of the cloud.

We will also define the local speed of sound $a$
\[ a^2 = \frac{1}{h} \left( \chi + \frac{p \kappa}{n^2} \right). \]
Symbols $\chi$ and $\kappa$ denote derivatives
\[ \chi = \left( \frac{\partial p}{\partial n}\right)_\epsilon, \;\;\; \kappa = \left( \frac{\partial p}{\partial \epsilon} \right)_n, \]
which have to be computed from an equation of state. The quantity $n$, the so-called baryonic density, is defined as a function assuring the conservation of the baryonic current $\nabla_\mu (n u^\mu) = 0$. Another quantity $h = (\rho + p)/n$ is the specific enthalpy. For barotropic equations of state the above formula can be reduced to $a^2 = d p / d \rho$.

By a quasi-stationary solution we will mean a solution for which neither the accretion rate defined as $\dot m = \partial_{t^\prime} m$ nor other hydrodynamic quantities, computed at a given areal radius, depend on time. In mathematical terms $\partial_{t^\prime} x = 0$, where $x = \dot m, U, p, \rho, \dots$ For the accretion case such an assumption can only be satisfied approximately, as we have already mentioned in the introductory section. Another assumption on the solution is that of asymptotic flatness of the space-time. The accretion cloud of our model is compact and has a finite mass. In the simplest version of the model such a cloud can be joined to the outside vacuum Schwarzschild space-time by a thin transient zone. Another possibility is to consider so-called ``quasi-stars'' \cite{begelman_et_al}. They consist of a spherically symmetric accretion cloud surrounded by a huge stellar-type atmosphere providing a reservoir of gas to feed the accretion process.

Under the aforementioned conditions, the Einstein equations and the equations of conservation of the energy-momentum tensor can be written as \cite{malec, kkmms, mach_malec}
\begin{eqnarray}
\frac{d}{d R} \ln \left( \frac{N}{k R} \right) = \frac{16 \pi}{k^2 R} (\rho + p), \;\;\;  U = \frac{C}{R^2 n}, \nonumber \\
N = \frac{B n}{\rho + p}, \;\;\; k R = 2 \sqrt{1 - \frac{2m}{R} + U^2}.
\label{steady_eqns}
\end{eqnarray}
Here $B$ and $C$ are integration constants. These equations are derived for a barotropic equation of state, i.e., one for which $p = p(\rho)$ or, equivalently, $p = p(n)$. From now on, we assume the polytropic equation of state $p = K n^\Gamma$.

The set of boundary conditions that we choose for above equations consists of the areal radius of the outer boundary of the cloud $R_\infty$, its total asymptotic mass $m_\mathrm{tot}$, the value of the local speed of sound at the outer boundary $a_\infty$, and the asymptotic baryonic density $n_\infty$. These conditions do not specify the solution completely. In order to do so one should also specify the asymptotic value of the velocity $U_\infty$. We will, however, adhere to a different, more standard approach. Namely, we will search for the transonic solutions --- those for which in outer regions of the cloud the flow is subsonic (the velocity of the fluid is smaller than the local sound velocity) and it becomes supersonic in the central parts of the cloud. Such a flow passes through the so-called sonic point, where $2U/(k R) = a$. Technically, this requirement means that one has to find a suitable value of $U_\infty$ for which the flow will be transonic.

There are two possibilities concerning the treatment of the central object. Suppose we attempt to integrate the equations starting from the outer boundary inward, and assume that a transonic solution is obtained. At a certain point the optical scalar $\theta_+ = k R/2 + U$ will vanish and that will correspond to the existence of an apparent horizon of a black hole. The areal radius of the apparent horizon will by further denoted as $R_\mathrm{BH}$. By the mass of the black hole corresponding to a given solution we will understand the quasi-local mass enclosed within the apparent horizon $m_\mathrm{BH} = m(R_\mathrm{BH})$. We will also define the mass of the fluid by $m_\mathrm{fluid} = m_\mathrm{tot} - m_\mathrm{BH}$.

Alternatively, we can stop the integration earlier and assume that at a given point a solid surface of the central object is met. This latter approach is used in the model of spherical accretion with radiation. It is assumed that the central object does not reflect the inflowing gas, but the corresponding energy is emitted in the form of radiation which then propagates through the accreting cloud \cite{karkowski_malec_roszkowski}.

\begin{figure}
\begin{center}
\includegraphics[width=100mm]{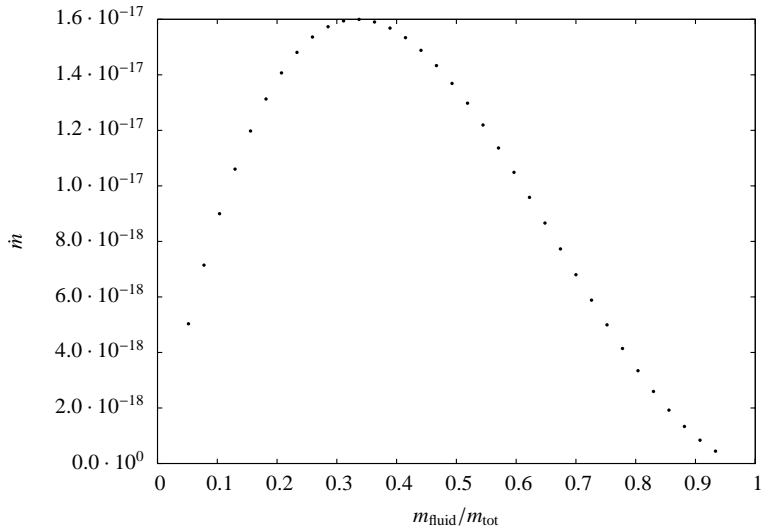}
\end{center}
\caption{The dependence of the accretion rate $\dot m$ on the $m_\mathrm{fluid}$. Here $m_\mathrm{tot} = 1$.}
\label{m_dot_graph}
\end{figure}

Fig.\,\ref{m_dot_graph} shows the dependence of the accretion rate on the ratio $m_\mathrm{fluid}/m_\mathrm{tot}$. Here all boundary parameters are kept fixed except for the asymptotic baryonic density $n_\infty$ (the ratio $m_\mathrm{fluid}/m_\mathrm{tot}$ scales linearly with $n_\infty$). In most cases there are two solutions corresponding to a given accretion rate. One, for which the mass of the central object is larger than the mass of the accreting fluid and the other in which the mass of the fluid constitutes most of the total mass of the accretion cloud. It can be shown analytically that the maximum of the accretion rate is achieved for $m_\mathrm{fluid}/m_\mathrm{tot} = 1/3$. Solutions for which the mass of the accreting fluid is small can be approximated by the well known test-fluid solutions. The other branch of massive solutions has been recently discovered in \cite{kkmms}.

\section{Stability}

\subsection{Numerical codes}

The stability of massive solutions has been investigated numerically using a general-relativistic hydrodynamical code. Some details concerning its construction can be found in \cite{mach_malec}.

The code is based on a modern high resolution shock capturing hydrodynamical scheme, where the (possibly discontinuous) solution of the equations of hydrodynamics is obtained by considering local Riemann problems that appear at interfaces between adjacent grid zones. In some respect, it is similar to a code constructed by Romero et al.\,\cite{romero_1996}. The implementation of the local Riemann solver is that of Donat and Maraquina\,\cite{donat_maraquina}. The reconstruction procedure required to obtain suitable Riemann states uses the ``minmod'' function of Van Leer\,\cite{van_leer}.

The code is formulated in the standard polar gauge with the metric of the form
\begin{equation}
\label{polar_metric}
d s^2 = - \alpha^2 d t^2 + X^2 d R^2 + R^2 \left( d \theta^2 + \sin^2 \theta d \phi^2 \right),
\end{equation}
where the lapse $\alpha$ and $X$ are functions of the areal radius $R$ and time $t$.

The essential point in the construction of the shock capturing scheme is the suitable, conservative formulation of the equations of hydrodynamics. Following certain standards established in numerical hydrodynamics community we use the Lorentz factor $W = \alpha u^t$ and the radial velocity $v^R = u^R/W$ (cf.\,\cite{banyuls_et_al}). Then, the equations of hydrodynamics, i.e., $\nabla_\mu \left( n u^\mu \right) = 0$ and $\nabla_\mu T^{\mu \nu} = 0$ can be written as
\[ \partial_t \mathbf q + \frac{1}{XR^2} \partial_R \left( \alpha X R^2 \mathbf F \right) = \alpha \mathbf \Sigma - 
\left( \partial_t \ln X \right) \mathbf q, \]
where $\mathbf q$ denotes a vector of conserved quantities
\[ \mathbf q = \left( D, S, \tau \right)^T = \left( n W, n h W^2 v_R, n W (h W - 1) - p \right)^T, \]
$\mathbf F$ is the flux vector
\[ \mathbf F = \left( n W v^R, n h W^2 v_R v^R + p, n W (h W - 1) v^R \right)^T, \]
and $\mathbf \Sigma$ denotes the source terms
\[ \mathbf \Sigma = \left( \begin{array}{c}
0 \\
\left( n h W^2 v_R v^R + p \right) \frac{\partial_R X}{X} - \left( n h W^2 - p \right) \frac{\partial_R \alpha}{\alpha} 
+ \frac{2 p}{R} \\
- n h W^2 v^R \frac{\partial_R \alpha}{\alpha} - \left( n h W^2 v_R v^R + p \right) \frac{\partial_t X}{\alpha X}
\end{array} \right). \]
Here $v_R$ is the covariant radial velocity $v_R = X^2 v^R$.

In the polar gauge, the Einstein equations reduce to just two partial differential equations that can be easily integrated by quadratures provided that the hydrodynamical source terms are already computed \cite{gourgoulhon}. The first of these equations can be easily expressed as
\[ \partial_R m = 4 \pi R^2 \left( n h W^2 - p \right) = 4 \pi R^2 (D + \tau), \]
where the quasi-local mass $m$ and the metric coefficient $X$ are related by
\begin{equation}
X = \frac{1}{\sqrt{ 1 - \frac{2m}{R}}}.
\label{X_function}
\end{equation}
The second one provides the radial derivative of the logarithm of the lapse
\begin{eqnarray*}
\partial_R \ln \alpha & = & X^2 \left( \frac{m}{R^2} + 4 \pi R \left( n h W^2 v_R v^R + p \right) \right) \\
& = & X^2 \left( \frac{m}{R^2} + 4 \pi R \left( S v^R + p \right) \right).
\end{eqnarray*}
Thus, at each time-step, basing on the known values of hydrodynamic quantities and metric functions for the previous time, one computes current values of hydrodynamic quantities and proceeds to integrate Einstein equations.

For dynamical calculations we use the perfect gas equation of state $p = (\Gamma - 1)n \epsilon$, where the specific internal density is defined by the relation $\rho = n + n\epsilon$. For smooth and isentropic flows the perfect gas equation of state reduces to the polytropic equation of state introduced in the preceding section. These two equations of state differ in results for discontinuous solutions.

The initial data for the dynamical code consists of five functions of areal radius $R$: $v^R$, $m$, $n$ and $\epsilon$. These can be obtained by numerical integration of equations (\ref{steady_eqns}) followed by a transition to polar coordinates. The latter is achieved using the following relations
\[ X = \left( 1 - \frac{2m}{R} \right)^{-\frac{1}{2}}, \;\;\; v^R = \frac{U}{\sqrt{1 + X^2 U^2}}. \]

The equations describing quasi-stationary flows were solved using \textsc{Daspk} --- a solver for a system of differential and algebraic equations developed by Petzold, Brown, Hindermarsh, and Li~\cite{brown_et_al}. In order to find the transonic solution we implemented a suitable bisection method taking into account that the transonic solution can usually be integrated toward much smaller values of $R$ than the other solutions.

At the inner boundary of the numerical grid the standard outflow conditions were assumed. This boundary was set outside the apparent horizon but always below the sonic point, so that the outflow conditions could be easily implemented (it is well known that satisfying such conditions in a subsonic region causes difficulties and one usually deals with some unwanted reflection from the grid boundary).

The outer boundary was kept fixed using the values obtained from the initial solution (inflow boundary). This condition cannot be easily relaxed; it is, for instance, also necessary for the stability of the test-fluid accretion.

The initial solution can be evolved for an arbitrarily long time without producing any noticeable changes. Thus, in order to investigate the stability of the flow, the initial data was perturbed in the velocity. The perturbation was chosen in the form of a bell-shape profile of a sine function restricted to one half of its period.

\subsection{Small perturbations}

\begin{figure}[t!]
\begin{center}
\includegraphics[width=100mm]{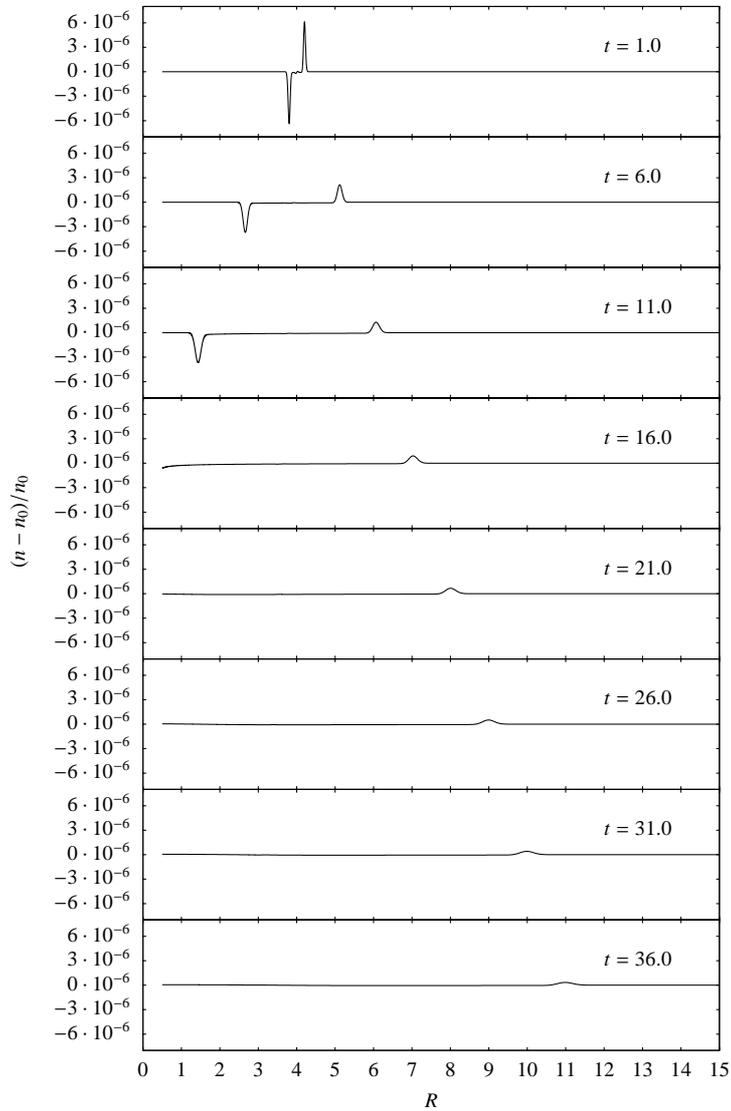}
\end{center}
\caption{Evolution of small perturbations. Here $n_0$ denotes the density in the unperturbed flow. The snapshots show the density contrast $(n - n_0)/n_0$ in the chronological order.}
\label{small_perturbations}
\end{figure}

Fig.\,\ref{small_perturbations} presents the evolution of a very small and compact perturbation. Here the initial quasi-stationary solution corresponds to the following parameters. The exponent in the polytropic equation of state was set to $\Gamma = 1.4$; the mass of the whole cloud $m_\mathrm{tot}$ was normalized to unity; the areal radius of the outer boundary of the cloud was equal $R_\infty = 10^6$; the asymptotic baryonic density was set to $n_\infty = 1.6 \cdot 10^{-19}$, and the asymptotic sound speed was such that $a_\infty^2 = 0.1$ (here and in all other results we have adopted the gravitational units where $c = G = 1$). For this solution the sonic point and the apparent horizon are located at $R_\ast = 0.82$ and $R_\mathrm{BH} = 0.34$ respectively. The mass of the fluid constitutes 83\% of the total mass. The accretion rate is equal $\dot m = 2.6 \cdot 10^{-18}$. This value is so small that the growth of the central mass can be completely neglected during the entire simulation.

The subsequent snapshots show the density contrast $(n - n_0)/n_0$, where $n_0$ corresponds to the baryonic density of the unperturbed flow (the unperturbed solution has to be evolved separately in order to subtract numerical noise that naturally appears even due to the interpolation of the initial data to the numerical grid of the dynamical code). The initial perturbation divides into two signals: one propagating toward the center and the other that propagates outward. The amplitudes of both signals decrease; the first signal eventually disappears in the inner boundary, while the second disperses until it is no longer distinguishable from the numerical noise.

\subsection{Strong perturbations}

\begin{figure}[t!]
\begin{center}
\includegraphics[width=100mm]{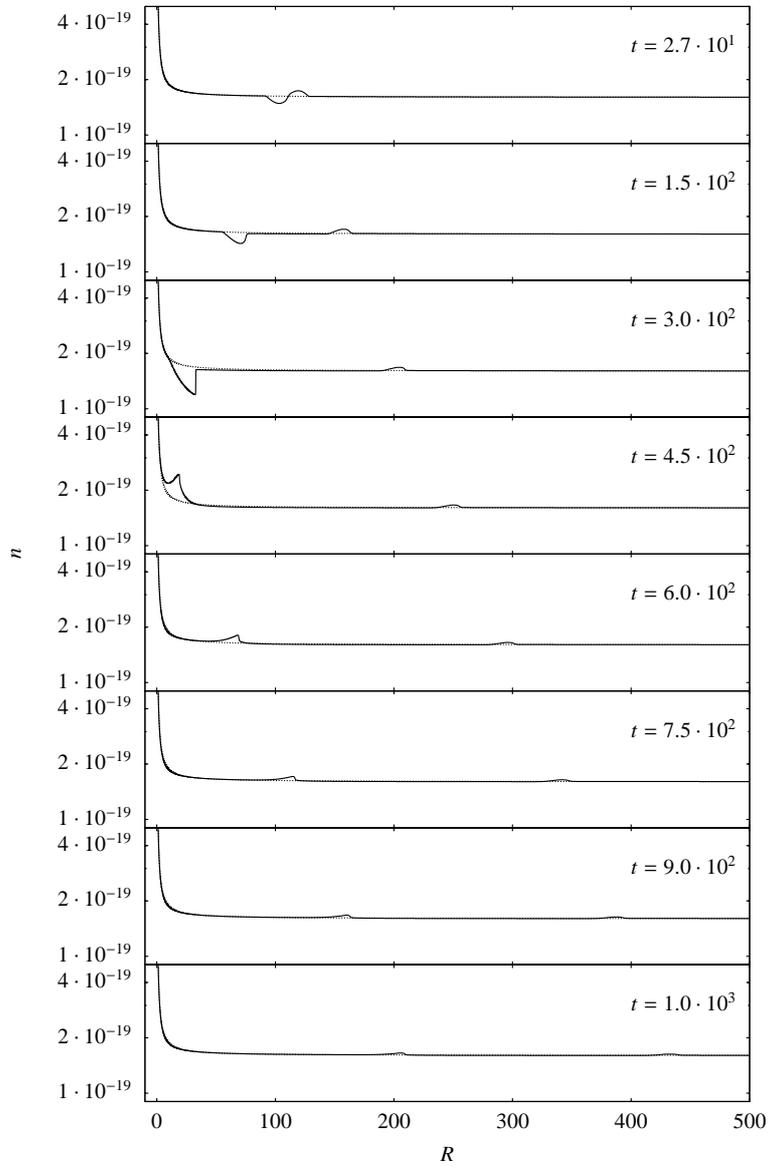}
\end{center}
\caption{Evolution of the perturbed density. The snapshots are placed in the chronological order. The profile corresponding to the unperturbed flow is depicted with a dotted line.}
\label{density_evolution}
\end{figure}

In the case of stronger perturbations the flow is still stable but the overall picture can look slightly different. Fig.\,\ref{density_evolution} shows the evolution of the baryonic density for such perturbations applied to the same initial solution as before. This time the incoming signal forms a shock wave and some part of it is reflected from the central regions of the accretion cloud. This is somewhat surprising. A small perturbation passing through the sonic point enters a zone in which the gas flows onto a center with a velocity larger than the local speed of sound, and thus it cannot be reflected outward; in order to make this possible such a signal would have to travel through the surrounding fluid with a supersonic velocity. Because of that fact a term ``sonic horizon'' has been coined to name the surface enclosing the supersonic region. This reasoning breaks in the nonlinear regime of strong perturbations. In this case the sonic point can move; new sonic points can be created; others can be destroyed. Moreover, the propagation velocity of shock waves that can eventually appear is not limited to the local speed of sound.

\begin{figure}[t!]
\begin{center}
\includegraphics[width=100mm]{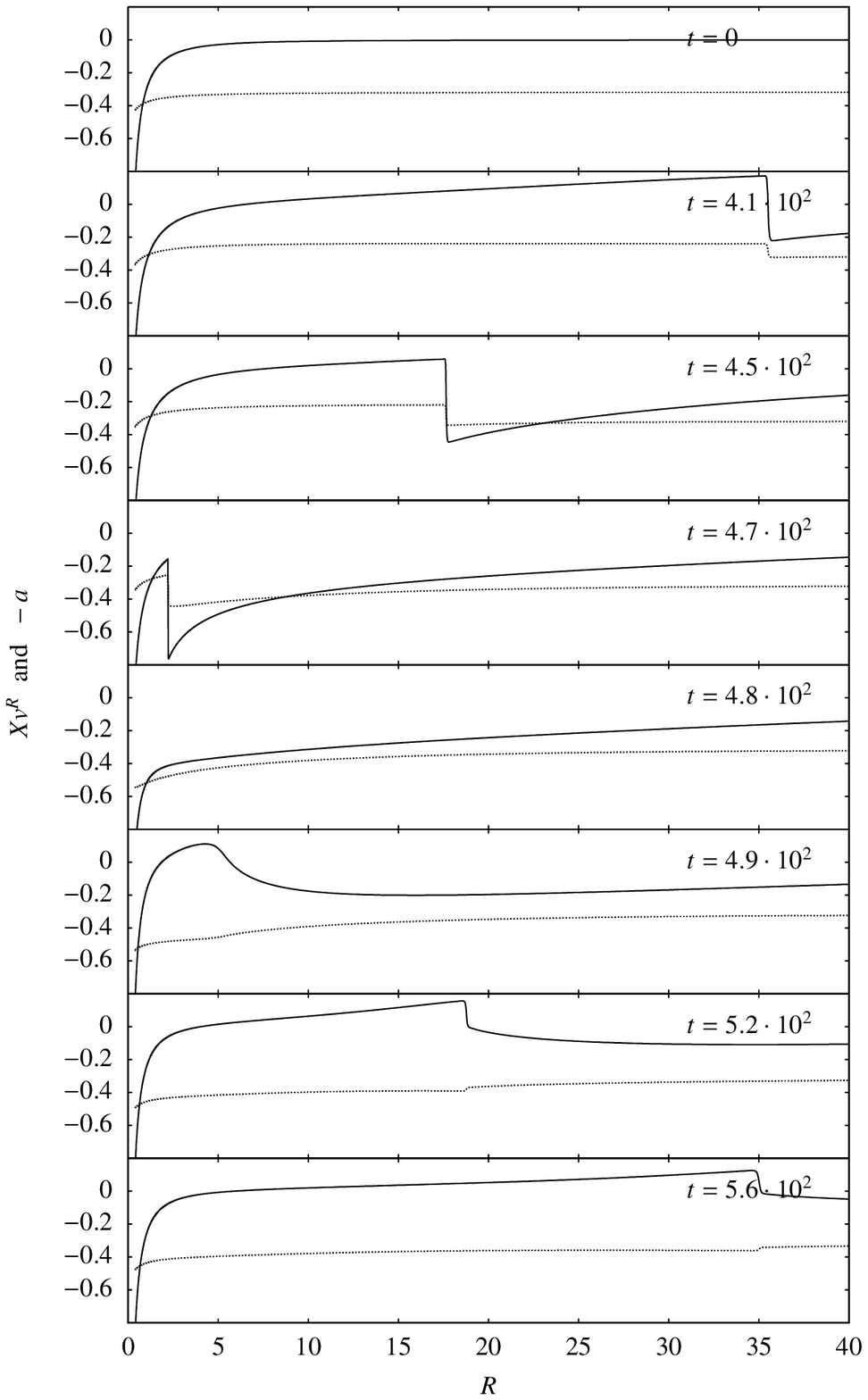}
\end{center}
\caption{Reflection of the incoming signal from the inner parts of the accretion cloud. The velocity $Xv^R$ is drawn with the solid line, while the dotted line depicts values of $-a$. Intersections of the two graphs correspond to the sonic points (see the discussion in text).}
\label{reflection}
\end{figure}
 
The reflection of a strong and wide perturbation is illustrated on Fig.\,\ref{reflection}. Here, for clarity reasons, the initial solution was perturbed slightly more than in the last case. The solid line shows the graph of $X v^R$, while the dotted line corresponds to $-a$. The condition for the existence of a sonic point expressed in terms used in our numerical code reads $|X v^R| = a$, i.e., it corresponds to the intersection of graphs of $Xv^R$ and $-a$. The incoming perturbation creates a new sonic point which starts to move toward the center of the cloud. At a certain stage the original sonic point is destroyed and, after relatively short time, it is replaced with the newly created one. In this process, the reflected signal is released and starts to propagate outward.

\begin{figure}[t!]
\begin{center}
\includegraphics[width=100mm]{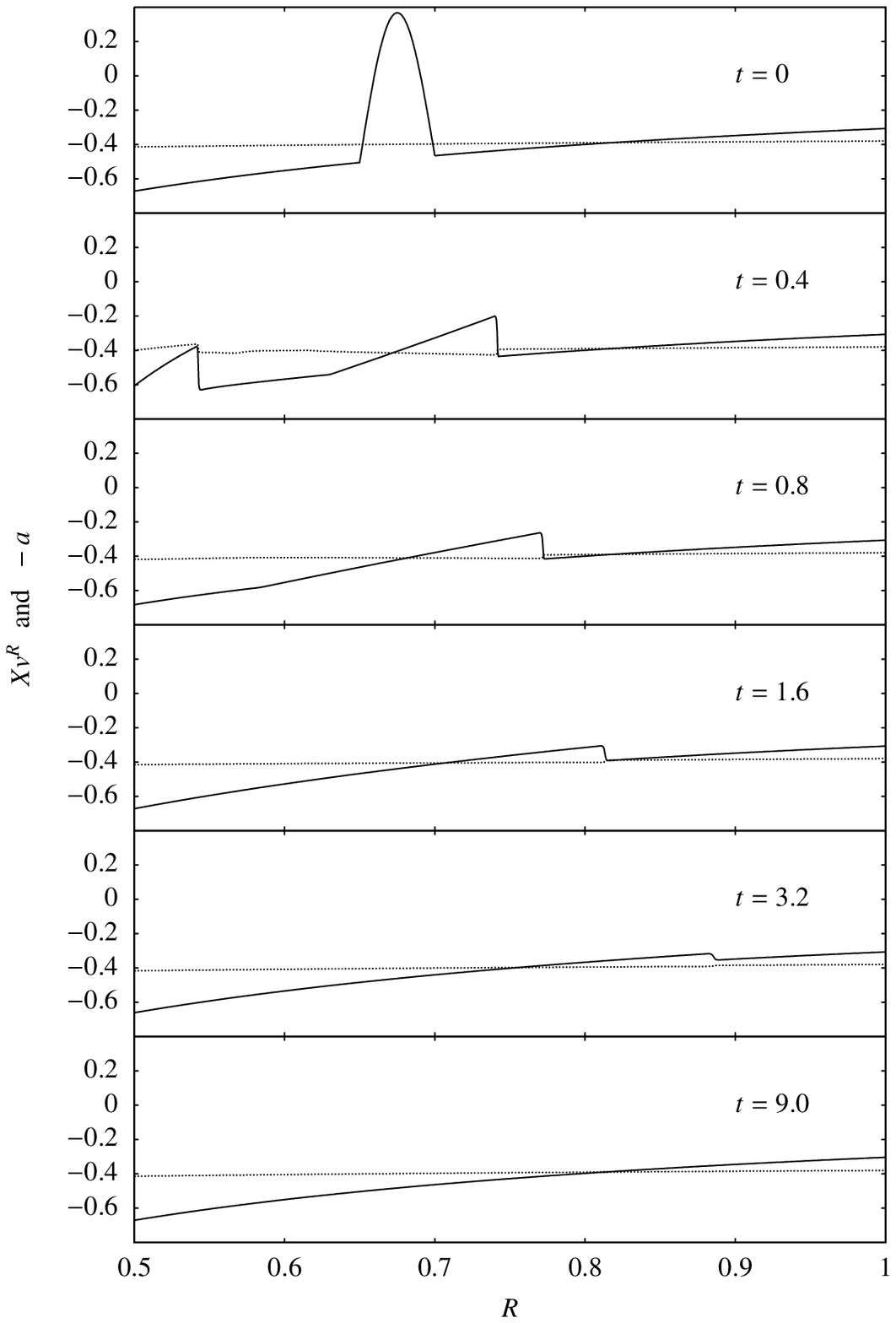}
\end{center}
\caption{A strong perturbation escapes form beneath the ``sonic horizon.'' Here, similarly to Fig.\,\ref{reflection}, the solid line is used for a graph of $X v^R$ and the dotted line for $-a$.}
\label{escape}
\end{figure}

Such behavior can be observed for perturbations whose support increases to a size larger than the size of the supersonic zone in the steady solution. We can, however, show that a strong perturbation which is initially located entirely in the supersonic zone can still escape from beneath the ``sonic horizon.'' Such case is depicted on Fig.\,\ref{escape}. Here, from an initially smooth data two shock waves are formed and the one which travels outward can eventually pass the original sonic point.

These results hint to the limited validity of the formal analogy between ``sonic horizons'' and event horizons known from general relativity (cf.\,\cite{das,dasgupta}). Contrary to the general-relativistic case the notion of the ``sonic horizon'' is restricted to the linear regime of small and compact perturbations.

Another observation concerning te case illustrated on Fig.\,\ref{escape} is that this time the perturbation traveling inward is strong but relatively narrow, and there is no reflection from the inner boundary.

\section{Conclusions}

In this paper we have discussed the stability of the general-relativistic spherically symmetric accretion onto compact objects. Calculations which take into account the self-gravitation of the accreting gas reveal the existence of two classes of weakly accreting steady solutions: one which can be well approximated by the test-fluid solutions and the other where the mass of the fluid is large. Our numerical results suggest that solutions belonging to both classes are stable against spherical perturbations (for the test-fluid solutions this can be proved analytically). The stability of solutions in the analogous Newtonian model has been reported in \cite{zakopane,mach_malec}. The Newtonian results presented in \cite{mach_malec} show the stability even for non spherical perturbations.

The advantage of the numerical approach is that we are no longer restricted to the linear regime of small perturbations. For strong self-gravitating perturbations the flow was stable in all examined cases. Investigation of the stability against strong perturbations has lead to a remarkable observation concerning the sonic points. In this case they are movable and a strong perturbation which propagates against the direction of the flow can easily escape from a supersonic region. This shows that the idea of the so-called ``sonic horizons'' makes only sense in the linear regime of small, purely acoustic signals.

Our analysis of the stability is restricted only to the accretion cloud. In all numerical simulations we had to ensure that the gas is delivered to the system at a small constant rate. To relax this assumption one could explicitly take into account some physical process responsible for such a ``feeding'' of the accretion cloud. One of the astrophysical possibilities is to consider the so-called ``quasi-stars'' described in \cite{begelman_et_al}. Another family of objects to which our analysis might be applied consists of Thorne and \.Zytkow stars \cite{thorne,thorne2}.

More realistic models of accretion should take into account the radiation emitted and transported through the accreting gas. Such models, which take into account the self-gravitation of the fluid, were recently analyzed in \cite{karkowski_malec_roszkowski}. In essential parts they agree with the results of our, purely hydrodynamical approach.

The model of self-gravitating accretion presented in this paper can be also used as a test problem for complex general-relativistic hydrodynamical numerical codes. Steady solutions are easy to be obtained; they are stable, and the proper treatment of self-gravity is necessary in order to preserve them unchanged in the dynamical code.

Another advantage of this model is that it allows for the investigation of the effects caused by self-gravity in a simple, yet not trivial scenario. Although the solutions cannot be written in a closed form, many of their properties can be obtained analytically (cf.\,\cite{kkmms,ladek}).

Acknowledgments. I wish to express my gratitude to Prof.\,Edward Malec for directing my attention to the studies of the stability of steady accretion and his enormous help in accomplishing this project. Some of the numerical computations reported in this paper were made at the Academic Computer Center Cyfronet, grant MNiSW/SGI3700/UJ/116/2007.


\begin{thebibliography}{aa}
\bibitem{banyuls_et_al} F. Banyuls, J.A. Font, J.$\mathrm{M^{\underline{a}}}$ Ib\'{a}\~{n}ez, J.$\mathrm{M^{\underline{a}}}$ Mart\'{\i}, J.A. Miralles, \textit{Astrophys. J.} \textbf{476}, 221 (1997).
\bibitem{begelman_et_al} M.C. Begelman, E.M. Rossi, Ph.J. Armitage, \textit{Mon. Not. R. Astron. Soc.} in press.
\bibitem{bondi} H. Bondi, \textit{Mon. Not. R. Astron. Soc.} \textbf{112}, 195 (1952).
\bibitem{brown_et_al} P.N. Brown, A.C. Hindmarsh, L.R. Petzold, \textit{SIAM J. Sci. Comput.} \textbf{15}, 1467 (1994).
\bibitem{das} T.K. Das, \textit{Class. Quant. Grav.} \textbf{22}, 2971 (2005).
\bibitem{dasgupta} S. Dasgupta, N. Bili\'{c}, T.K. Das, \textit{Gen. Rel. Grav.} \textbf{37}, 1877 (2005).
\bibitem{donat_maraquina} R. Donat, A. Maraquina, \textit{J. Comput. Phys.} \textbf{125}, 42 (1996).
\bibitem{gourgoulhon} E. Gourgoulhon, \textit{Astron. Astrophys.} \textbf{252}, 651 (1991).
\bibitem{kkmms} J. Karkowski, B. Kinasiewicz, P. Mach, E. Malec, Z. \'{S}wierczy\'{n}ski, \textit{Phys. Rev.} \textbf{D73}, 021503(R) (2006).
\bibitem{karkowski_malec_roszkowski} J. Karkowski, E. Malec, K. Roszkowski, \textit{Astron. Astrophys.} \textbf{479}, No. 1, 167(2008)
\bibitem{ladek} B. Kinasiewicz, P. Mach, E. Malec, \textit{International Journal of Geometric Methods in Modern Physics} \textbf{4}, 197 (2007).
\bibitem{van_leer} B. Van Leer, \textit{J. Comput. Phys.} \textbf{32}, 101 (1979).
\bibitem{zakopane} P. Mach, \textit{Acta Phys. Pol.} \textbf{B38}, 3935 (2007).
\bibitem{mach_malec} P. Mach, E. Malec, \textit{Phys. Rev.} \textbf{D78}, 124016 (2008).
\bibitem{malec} E. Malec, \textit{Phys. Rev.} \textbf{D60}, 104043 (1999).
\bibitem{michel}F.C. Michel, \textit{Astrophys. Space Sci.} \textbf{15}, 153 (1972).
\bibitem{moncrief} V. Moncrief, \textit{Astrophys. J.} \textbf{235}, 1038 (1980).
\bibitem{romero_1996} J.V. Romero, J.$\mathrm{M^{\underline{a}}}$ Ib\'{a}\~{n}ez, J.$\mathrm{M^{\underline{a}}}$ Mart\'{\i}, J. Miralles, \textit{Astrophys. J.} \textbf{462}, 839 (1996).
\bibitem{thorne} K.S. Thorne, A.N. \.{Z}ytkow, \textit{Astrophys. J.} \textbf{199}, L19 (1975).
\bibitem{thorne2} K.S. Thorne, A.N. \.{Z}ytkow, \textit{Astrophys. J.} \textbf{212}, 832 (1977).
\end{thebibliography}
\end{document}